# Space-time adaptive control of femtosecond pulses amplified in a multimode fiber

**Raphael Florentin, Vincent Kermene, Agnes Desfarges-Berthelemot, Alain Barthelemy,** *

Université de Limoges, XLIM, UMR CNRS 7252, 123 Avenue A. Thomas, 87060 Limoges, France

*Corresponding author: alain.barthelemy@xlim.fr

**Ultrashort light pulse transport and amplification in a 1.3 m long step-index multimode fiber with gain and with weak coupling has been investigated. An adaptive shaping of the input wavefront, only based on the output intensity pattern, has led to an amplified pulse focused both in space (1/32) and in time (1/10) despite a strong modal group delay dispersion. Optimization of the input owing to the two-photon detection of the amplified signal permitted to excite the fastest and more intense principal mode of the fiber and to get an output pulse duration limited by group velocity dispersion.**

## 1. Introduction

Multimode optical fibers (MMFs) are currently intensively revisited for their various potential applications in optical communications [1], in endoscopic imaging [2,3], in high resolution spectral analysis [4], in quantum processing [5], etc. Because of mode dispersion and of random mode coupling occurring in long or stressed multimode fibers, propagation in MMF shares some common features with propagation through scattering media [6]. It is well known for instance that a Gaussian laser beam injected in a multimode fiber is generally scrambled in its intensity profile and delivered as a random speckle at the fiber output. Capitalizing on the techniques previously developed for spatial shaping of the waves transmitted through scattering media [6,7], it was shown that the laser field delivered by MMF can be controlled in its transverse distribution through wavefront shaping of the input field. The highest level of control is offered by the transmission matrix approach [8] and by digital holography phase conjugation [9]. Both methods assume a linear propagation regime, a stable fiber and require a calibration step. In particular, they are not able to adapt to perturbations of the fiber without a new calibration. Adaptive approach, for which the patterning of the wavefront by a spatial light modulator (SLM) is performed through the optimization of a cost function [10], appears to be well suited to simple spatial shaping almost in real time in a non-protected environment. This is also a technique which was recently demonstrated to be compatible with nonlinear propagation regime such as in free space Kerr liquid [11] and in amplifying multimode fiber with gain saturation [12]. The three kinds of methods permit beam focusing and imaging through MMF. Most of the papers published up to now investigated continuous wave light transport through MMF. Recently space time analysis and control of light pulses delivered by MMF became a topic of great interest. As it occurs for the transmitted laser field in the space domain, an ultrashort pulse delivered by a MMF is usually scrambled in profile and stretched in the time domain. This is a consequence of both mode dispersion (phase and group delays) and mode coupling. Spatiotemporal structures at the output of multimode fibers were characterized by time sampled interferometric measurement [13] or by spectrally resolved interferometric measurement, either with short pulse excitation [14] or with a CW tunable laser source [15]. The more complete characterization of space time distortions due to propagation in an MMF is provided by measurement of the MMF transmission matrices for a large set of input wavelengths [15]. It gives an extended 3D transmission matrix from which one can derive the spatiotemporal input field required to get a desired spatiotemporal output. Furthermore, these data permit computation of the time delay matrix and determination of its associated eigenstates, the principal modes (PMs) [16]. The PMs, corresponding to Wigner-Smith states, are specific complex field patterns which are transmitted through the MMF with a strongly reduced impact of mode group delay dispersion (to first order). Without coupling, the PMs simply correspond to the MMF eigenmodes. PMs were studied recently in the weak and strong mode coupling regimes [17]. However, the use of a PM to deliver a short pulse at the end of a MMF has not been yet reported. Adaptive wavefront shaping was also used to perform some spatiotemporal control at the output of a MMF. It was shown for example that the optimization of the two photon fluorescence signal from a uniform target covering the whole MMF core may lead to a sharp focus on the fiber output face [18]. This is a consequence of the fact that a two photon nonlinear process depends on the local and instantaneous intensity. Most of the published literature deals with passive optical fibers in linear propagation regime. It is only recently that nonlinear MMF and MMF with gain have been considered in combination with transverse phase control for adaptive frequency generation [19] or adaptive spatial shaping [12]. It was shown in particular with a CW signal that the adaptive method achieves an efficient focusing of the amplified output even in gain saturation regime with the onset of gain competition among the fiber modes.

In this letter, we report new results on the various space time shaping one can perform at the output of an Yb-doped multimode fiber amplifier thanks to an adaptive pure spatial phase patterning of the input ultrashort pulse. We investigated the pulse compression associated with the time-integrated output pattern shaping in a single focus. The conditions leading to a short amplified pulse shaped in a single spot were identified. Then we studied the output pattern shaping associated with an optimization of the space and time integrated squared intensity. Such an optimization could give rise to the direct observation of the fiber principal mode of widest spectral bandwidth.

## 2. Space-time shaping into a short pulse focus

The experimental set-up schematically shown on Fig.1 is extremely simple. It consists first of a mode-locked Yb:KGW laser (Mikan, Amplitude System) delivering pulses of $\tau$ =244 fs duration at 38 MHz repetition rate and 1029 nm wavelength. The laser beam was magnified by a telescope so that its center part covered the deformable segmented mirror (DSM, kiloDM Boston Micromachines) with an almost uniform intensity. The reflected beam with a shaped wavefront was imaged onto the fiber input face by a second demagnifying telescope, to match the fiber core of 90 µm diameter. The amplifying fiber from Nufern exhibits a step index refractive index profile with an index difference of 3.8 $10^{-3}$ and it is uniformly Yb-doped in the core. The first cladding for guiding the pump radiation has an octagonal shape with a width of 400 µm and a numerical aperture of 0.46 (at 5%), giving a pump absorption of 11.5 dB/m at 976nm. We have used a 1.3 m long MMF piece. It was laid on the table and bent between its two holders with perpendicular orientation, making a quarter circle of 15 cm radius. The amplifying MMF was backward pumped through a dichroic mirror (D) by fiber coupled laser diodes at 976 nm.

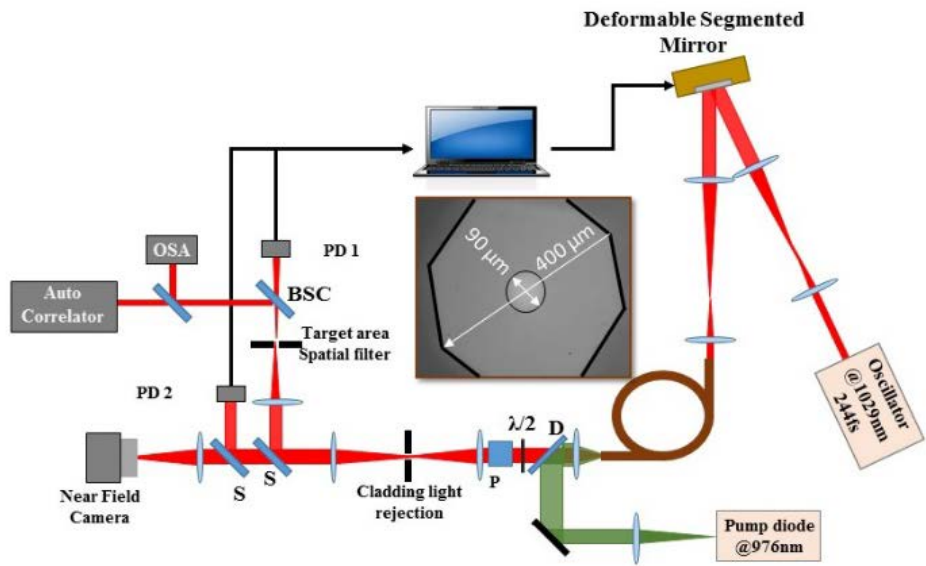

**Figure 1. Experimental set-up designed for space time shaping at the output of the Ytterbium doped multimode fiber amplifier (fiber section in inset).**

The pump absorption was about 95% in the fiber sample and the residual pump power exiting the fiber was blocked by a spectral filter (not shown) to avoid damages on the DSM. The pattern of the amplified laser wave on the MMF output, was imaged onto a screen where a pinhole served to filter out the power in the desired disc target before detection by a standard silicon photodiode (PD 1) (Thorlabs DET10A/M). The electric signal delivered by the photodiode was feeding the optimization routine on the computer which commanded the height of the different mirror segments. A polarizer P selected one vector component of the amplified laser field. Beam splitters (S, BSC) diverted a fraction of the output for beam observation and recording by a camera as well as for pulse measurement (SHG background free autocorrelation). The chromatic dispersion length is here $L_D$ =1.8 m. The group velocity was computed for the different linearly polarized modes of the fiber (104 LP modes for one polarization) giving the graph illustrated on Fig.2. The group delay standard deviation amounts to $\sigma_{GD}$=2.90 ps/m giving a modal dispersion length of $L_{MD}= \tau/\sigma_{GD}$ =8.41 cm.

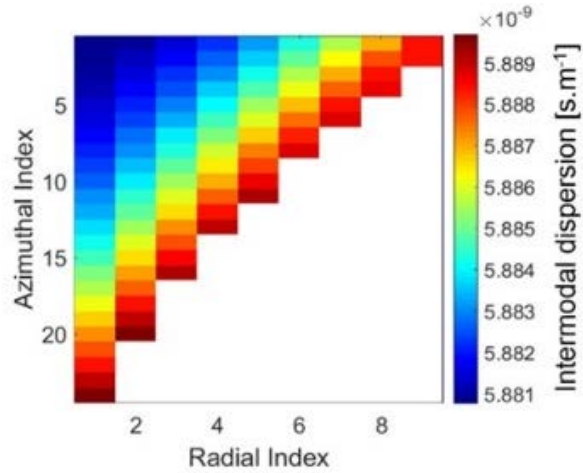

**Figure 2. Group delay of LP$_{mn}$ modes of the MMF.**

Therefore the propagation regime in the gain fiber of physical length L corresponded to the case L/L$_{MD}$>>1, where the laser pulse distortion was ruled out by the modes group velocity. When a random phase pattern was set on the DSM, we measured at the MMF output the autocorrelation trace shown in Fig.3-a, which is very close to the trace expected from numerical simulation, especially in its central part. It corresponds to a long pulse lasting 12 ps with a more powerful part having a ~7 ps FWHMI duration (~2.= $\sigma_{GD}$.L) in agreement with the transmission of the initial pulse split on many modes. This large spreading in time gave in turn a mostly incoherent summation of the mode transverse field at the fiber output which resulted in a speckled output pattern with low modulation contrast. In other words it means that the spectral correlation bandwidth of the MMF is significantly smaller (~1/15) than the input laser bandwidth. A typical recording is given in Fig.3-b.

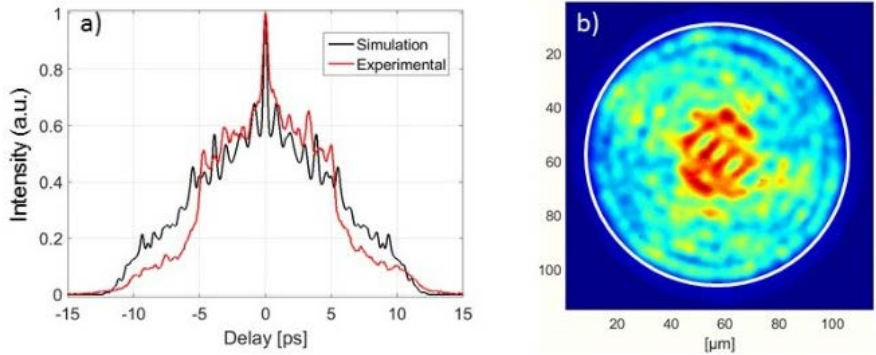

**Figure 3. Typical autocorrelation trace (a) of the pulse obtained with a random input wavefront and intensity distribution (b) at the output of the MMF. The white circle shows core-clad limit.**

We know from previous works carried out with CW laser that a local intensity enhancement and an output beam shaping in a single focus can be achieved by an adaptive structuration of the input wavefront. This is possible because of the coherent combining of the output modes and because of their linear mapping to the modes excited at the input. One can therefore wonder whether the adaptive approach for beam focusing through MMF can still be effective in short pulse regime with strong modal dispersion. To answer that question we used the same technique we have exploited in CW regime. In the set-up of Fig.1, for optimization of the DSM surface we used the time integrated power measured by the photodiode PD 1 after spatial filtering of the fiber output pattern by a hard circular aperture representing the surface and location of the desired spot. The optimization of the wavefront was still based on three states phase-shifting interferometry routine [12] and lasted about 2 seconds. For small size focus target, of the order of the smallest speckle grain $\lambda$/(2.NA)=5 µm, the wavefront optimization led to a sharp focus accompanied by a significant background in the rest of the core (Fig. 4(a)). The corresponding autocorrelation trace covered a time span of 6-8 ps with a profile indicating that the output signal is formed by a train of four main pulses (Fig. 4(b)). Therefore, the optimized pattern again resulted from a mixture of coherent and incoherent superposition of a number of modes, mainly those with a zero azimuthal number. For larger diameters of the focus target, the wavefront shaping still led to a clean bell-shaped focus carrying a fraction of the total power which increased according to the target size and then reached a plateau.

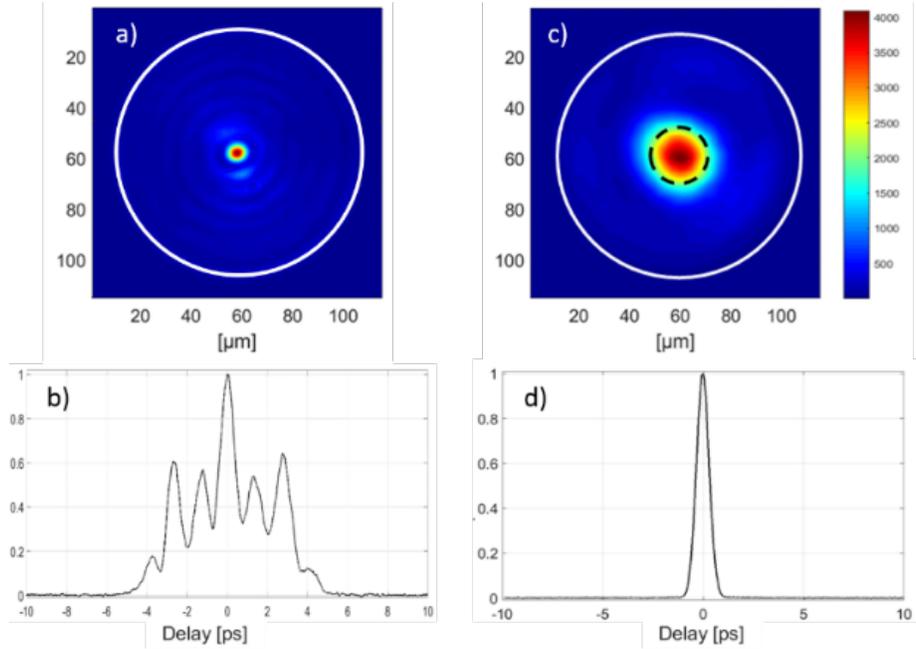

**Figure 4.** Output patterns (a) and (c) and corresponding autocorrelation traces (b) and (d) recorded after wavefront shaping. For figures (a) and (b) (resp. (c) and (d)), optimization of the input phase shape was made with a 5 µm (resp. 20 µm) diameter focus target. In Fig. (c), small black dashed circle denotes the target limit.

For a target disc of 20 µm diameter or larger, the time integrated image of the amplified spot on the fiber output face was of high intensity contrast (4096:6) with respect to the average background in the rest of the core (see Fig.4(c)) similarly to what was obtained with a CW input signal. The intensity enhancement at the focus was 32 by reference to the average intensity measured at the output for a random input wavefront. The image was recorded for a gain of 14dB. The transverse profile of the shaped focus was different from the one of the fiber fundamental mode and demonstrates that it did not resulted from the simple selection of the lowest order mode. This is consistent with our modeling of the whole experiment. At the same time the autocorrelation traces shrank significantly from ~6 ps down to ~760 fs FWHMI when the focus diameter was broadened from 5 µm to 20 µm (see Fig 4(d)). The autocorrelation remained short and clean without pedestal for focus up to 38 µm, the maximum value we have tested. So, the wavefront shaping for a pure spatial focusing of the average power delivered by the MMF amplifier, performed simultaneously a tenfold compression of the output pulse (at least). This is consistent with our modeling of the multimode propagation coupled with the optimization process. The shaping therefore achieved an enhancement in peak intensity greater than 320. Such a space-time coupling is not a surprise and the behavior is similar to the one observed when focusing ultrashort pulses through disordered media where the differential phase delays are due to multiple scattering rather than from mode dispersion and coupling [20]. It was also possible to move the desired focus spot on any position inside the fiber core. For a focusing close to the fiber core/cladding boundary, when shaping a spot with a size equal or greater than 20 µm, we got a slightly better pulse compression effect with an autocorrelation duration reduced to 535 fs (Fig. 5). Pulse compression as well as amplifier gain were not significantly altered by the beam shaping. For reference with a random pattern, we measured a gain of 14.3 dB which evolved to 14.9 dB when the amplified beam was focused on the center of the core and to 14.2 dB when it was focused on the core edge. This is consistent with the fact that there is here a very weak modal gain dispersion (weak gain drop for the highest order modes) since the core is uniformly doped in active ions. Comparison of the space-time shaping performances with and without gain gave close results both in terms of focusing quality and in terms of pulse compression. That was expected based on the fact that the gain is almost uniform in the modal population and that the amplifier was operated in the weak saturation regime.

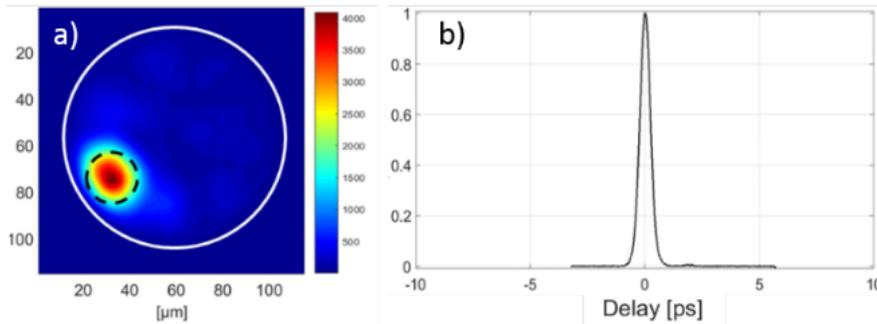

**Figure 5.** Output pattern (a) and corresponding autocorrelation traces (b) recorded after wavefront shaping. Optimization was performed with a 20 µm diameter focus target located on the black dashed circle.

## 3. Optimization for selection of a principal mode

The output field of a MMF fiber can be expanded into principal modes (PM) of the time delay (Wigner-Smith) matrix which form an orthogonal basis well suited to space-time analysis of the multimode propagation of short pulses. It is known from the pioneering works on MMF reported in [21] and in [22] that, in the weak coupling regime, the PMs with the broadest spectral correlation bandwidth are the first and last PM's of the time delay matrix, sorted in increasing time delay. The two spatial PMs are associated with the shortest pulses, and hence the highest peak powers, that can be delivered by the MMF. In the case of weak coupling, [17] shows that the fastest PMs are composed of low order fiber eigenmodes and are characterized by few wide speckle grains close to the core center. We modeled the propagation through the MMF amplifier according to the concatenated sections approach [17,21] with ten straight multimode fiber sections connected by unitary random coupling matrices. Considering a weak coupling regime and a slightly weaker linear gain for the high order modes (proportional to the overlap integral between the mode intensity and the doped area) we obtained a typical transmission matrix illustrated in Fig. 6(a). The corresponding PMs were computed and their features were found to be similar to the ones of [17] and [21]. For example the fastest PM output pattern is given in Fig.6-b. PMs of intermediate delay cover entirely the core cross-section and appear as complex speckle pattern with small scale modulations (see Fig. 6(c)). The figures are consistent with the fact they are made up of the overlap of a large number of eigenmodes. Consequently such PMs exhibit a small spectral correlation bandwidth. The slowest PMs are also composed of a reduced number of fiber eigenmodes (~15) but of high orders so that the PM pattern covers a large part of the core area (see Fig. 6(d)).

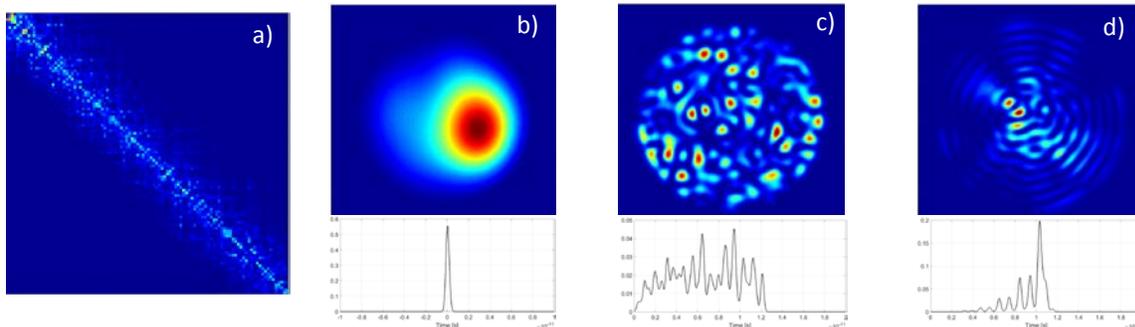

**Figure 6. (a) Transmission matrix of a MMF amplifier with linear gain and weak coupling used in the simulation, (b) principal mode with the smallest group time delay (PM#1), principal mode with an intermediate delay (PM#33), and slowest principal mode (PM#104). Corresponding pulse profiles in lower insets.**

A good way to preserve a short duration after amplification in a MMF amplifier of a short laser pulse would be to shape it at input in a broadband PM. This would require the measurement of the MMF transmission matrix at different wavelengths to compute the Wigner Smith matrix. We proceeded here in a different way and we used the previous adaptive wavefront shaping set-up to shape the signal beam input into a PM, in order to get the shortest amplified pulse which can be delivered by the MMF. For that goal, we chose as a feedback signal for the optimization process a peak power dependent measurement based on a two photon absorption (TPA) photodetector (PD) without performing any spatial filtering. We first computed the TPA-PD signal that would be generated by a spatial input PM excited with a pulse of 240 fs duration similar to the one of our laser. This duration corresponds to a bandwidth close to the fastest PM spectral correlation bandwidth. For each PM input, its output 3D space-time intensity distribution was computed and the corresponding nonlinear signal value was derived. The result plot on Fig.7 indicates that the fastest PM should produce the stronger nonlinear signal. This is in contrast to the case investigated in [18] where the same kind of optimization gave rise to a sharp focus because the mode group delay dispersion was of low impact in the considered experimental situation. On the experimental set-up we replaced the standard PD by a TPA-PD (Hamamatsu G116) and we removed the aperture which performed the selection of fiber core area. The same optimization as in the previous cases was achieved and yield the time integrated spatial shape shown in Fig.8. It is very close to the one derived from the modelling (see Fig. 6(b)). The fact that the output pattern after optimization differed from a pure fiber eigenmode attest the presence of coupling along the light path, even if weak, similarly to what was accounted for in the model. In addition the corresponding autocorrelation trace was measured to be shaped in a single peak with a width of only 518 fs (FWHMl). It is again in very good agreement with the result of the model and the pulse duration corresponds approximately to the limit fixed by the chromatic group velocity dispersion for a 240 fs input pulse, meaning that the modal dispersion was almost completely overcome.

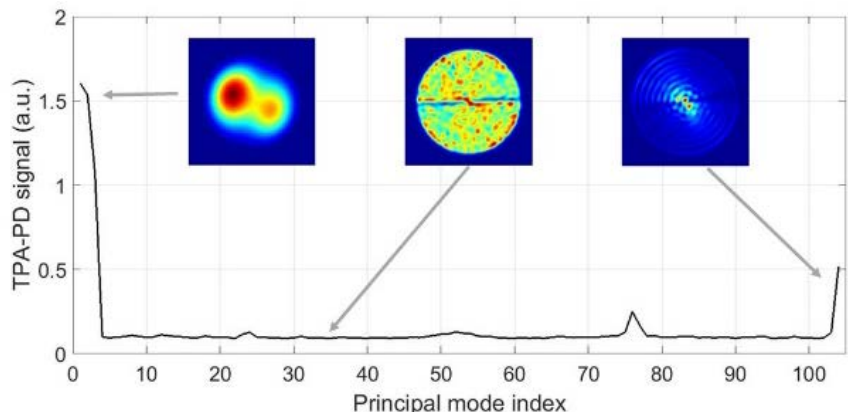

**Figure 7. Plot of the computed TPA-PD signal owing to the principal mode launched in the MMF amplifier. Insets show some time integrated output patterns for a 240 fs input pulse with transverse shape corresponding to the quickest, intermediate and slowest PMs.**

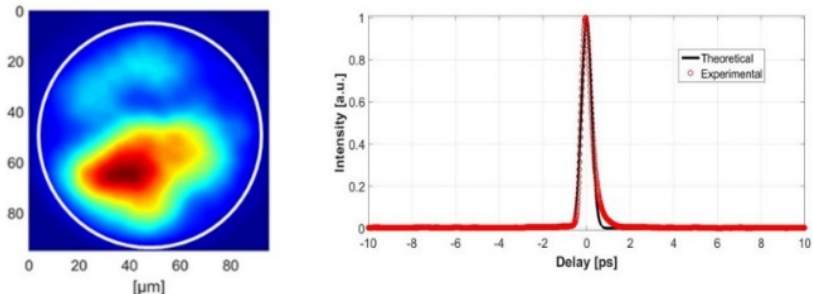

Figure 8. Near field intensity of the MMF output with gain after optimization of the TPA-PD signal, on the left. The experimental figure is close to that of the expected PM. Corresponding autocorrelation traces of the amplified pulses (red circles: measurements, and black line: theoretical PM trace) on the right.

## 4. Conclusion

Adaptive control of the spatiotemporal profile of a femtosecond pulse amplified in a MMF was investigated in the weak mode coupling regime and with a strong impact of modal group delay dispersion. Input wavefront shaping for pure spatial focusing of the amplified field was achieved and the simultaneous compression of the output pulse was measured. Both space and time shaping was shown to depend on the size of the desired focus as well as on its position in the core cross-section. The gain was weakly impacted by the shaping (+/- 0.3 dB) but a strong saturation regime was not reached (< 2 dB of gain reduction). Our observations were in good agreement with the results of the modeling of the whole process including MMF propagation and DSM profile optimization. In experiments we got amplified pulses of duration down to 357-378 fs shaped in a single spot with nearly Fourier transform limited divergence. Although we have reached, on the MMF output face, very high laser field intensity, the effective area of the guided radiation all along its path down the fiber is broad so that we did not measure any spectral broadening by Kerr effect even when we pumped up the average output power to 20 W where the output peak power was close to 1 MW. In a second step we looked for the MMF excitation yielding the strongest two-photon detection at the output. As a result, the adaptive shaping led to very short amplified pulses (354 fs duration, the shortest we measured) and to a beam profile made up of a few low order modes in good agreement with the excitation of the fastest principal mode of the MMF with linear gain. The modal group delay dispersion was fully compensated in that latter situation. So, adaptive wavefront shaping techniques, which do not require a calibration step, appear to be well suited to control the amplification of ultrashort light pulses in a MMF with gain without significant degradation of their space time distribution.

**Funding**. Agence Nationale de la Recherche, ANR-14-CE26-0035-01, POMAD project.

**Acknowledgment**. We thank Prof. H Cao for stimulating and enlightening discussions.